# Understanding Interface Stability in $RE$Ni$_2$/Ni through First-Principles Calculations

Yuta Yahagi[1,2]*, and Yumi Katasho[2]*

[1]*NEC Corporation, Minato-ku, Tokyo 108-8001, Japan*

[2]*National Institute of Advanced Industrial Science and Technology, Tsukuba, Ibaraki 305-8569, Japan*

*E-mail: yuta-yahagi@nec.com; katasho.yumi@aist.go.jp

Crystallographic orientation analysis revealed that DyNi$_2$ grew epitaxially on Ni, whereas NdNi$_2$ does not. To elucidate the microscopic origin of this contrasting behavior, we constructed atomistic models of Ni/Rare-earth ($RE$)Ni$_2$ interfaces with well-defined crystallographic alignment and performed first-principles calculations based on density functional theory (DFT).

The computed interfacial energies exhibit a clear correlation with lattice mismatch: larger mismatch leads to higher interfacial energy and reduced interface stability. Consequently, Ni/DyNi$_2$ exhibits a significantly lower interfacial energy than Ni/NdNi$_2$, consistent with experimental observations.

A comparison between interfacial and strain energies for Ni/$RE$Ni$_2$ ($RE$ = Sc, Y, Nd, Gd, Dy, and Lu) reveals that the elemental dependence of interfacial stability is dominated by elastic strain rather than chemical bonding. Based on this insight, we developed a simple regression model using the absolute lattice mismatch as a descriptor, enabling qualitative predictions of stability for Ni/$RE$Ni$_2$ interfaces with RE other than those examined in DFT.

# 1. Introduction

Rare-earth (RE) elements are indispensable for high-performance magnets, batteries, and other advanced technologies. Efficient RE recycling is therefore a key technological challenge for ensuring sustainable energy systems and resource security. A major challenge lies in separating individual RE elements from a mixture of multiple RE metals, as they exhibit highly similar chemical properties. One effective technique for RE separation is the selective alloy diaphragm method, which enables the separation of Dy and Nd in molten salt by controlling the electrochemical potentials of the alloy diaphragm and electrodes.[1)–3)]

This recycling process relies on the rapid electrochemical formation of RE–iron-group alloys in molten salt, which has been extensively investigated through experimental studies, including *in situ* methods.[4)–12)] At temperatures around 773 K, the electrochemical alloying of Dy and Ni proceeds substantially faster than thermal diffusion.[7)] The $RE$Ni$_2$ phase, in particular, adopts a cubic Laves structure and exhibits an anomalously high formation rate compared with other intermetallic compounds. Electrochemical alloying in LiCl–KCl–RECl$_3$ melts has been reported for a wide range of $RE$ elements, including Ni–Dy[6)–9)] and Ni–Nd[10)–12)], as well as Ni–Y[13)], Ni–La[14)], Ni–Ce[15)], Ni–Pr[16),17)], Ni–Sm[18)], Ni–Tb[19),20)], Ni–Yb[21)] and Ni–Lu[22)].

Recent studies revealed a striking contrast in growth behavior: DyNi$_2$ grows epitaxially on Ni (Figure 1(a))[9)], whereas NdNi$_2$ does not (Figure 1(b)), despite its confirmed formation by X-ray diffraction[12)]. Epitaxial DyNi$_2$ forms on both polycrystalline and single-crystal Ni electrodes.[9)] When the Ni substrate contains grain sizes of approximately 15 μm, the DyNi$_2$ alloy phase exhibits comparable grain sizes (approximately 15 μm)[9)], while NdNi$_2$ forms much smaller grains (<3 μm)[12)].

Although Dy and Nd share similar chemical properties as belonging to REs, their markedly different interfacial growth behaviors suggest a critical underlying physical mechanism. To clarify the origin of this difference, we performed first-principles calculations focusing on the interface stability of Ni/$RE$Ni$_2$ systems. Due to computational constraints, explicit calculations were conducted for $RE$=Sc, Y, Nd, Gd, Dy, and Lu.

We first constructed atomistic models of Ni/$RE$Ni$_2$ interfaces. We then evaluated their energetic stability density functional theory (DFT) calculations combined with equation-of-state fitting and explicit interfacial energy calculations. Finally, we discuss the observed trends in interface stability and predict interfacial energies for other $RE$ elements. Temperature effects and molten salt environments were neglected in this study, and we focused exclusively on the intrinsic interfacial energy.

## 2. Theoretical methods
## 2.1. Interface model

We constructed atomistic models of Ni/$RE$Ni$_2$ interfaces using a systematic procedure designed to suppress artificial electric fields, eliminate spurious surface contributions under periodic boundary conditions, and enable an unambiguous evaluation of interfacial energetics after structural relaxation.

(i) **Choice of base crystals.**

$RE$Ni$_2$ is modeled in the cubic Laves phase, and Ni in the fcc structure, as shown in Figure 2(a) and (b)[23]. Motivated by the approximate relation $a_{RENi_2} \approx 2a_{Ni}$, a $(2 \times 2 \times 2)$ supercell was used as the base structure of Ni.

(ii) **Nonpolar slab cleavage.**

Stoichiometric, nonpolar (001) slabs were cleaved from each base crystal to suppress macroscopic dipoles perpendicular to the interface[24),25)].

(iii) **Hetero-stacking with initial interfacial registry.**

$RE$Ni$_2$ and Ni slabs were placed in contact along the [001] direction to construct a coherent superlattice. The initial interface spacing was chosen to preserve the nearest-neighbor $RE$-Ni distance in bulk $RE$Ni$_2$, ensuring an unbiased initial registry. the lateral lattice constants were fixed to $2a_{Ni}$.

(iv) **Arranging identical interfaces.**

The number of Ni layers was adjusted to create crystallographically equivalent interfaces at the top and bottom of the supercell. This guarantees that the total interfacial contribution is exactly twofold, enabling the interfacial energy to be extracted without ambiguity (Section 2.2).

(v) **Packing atoms into voids.**

To emulate the energetic preference for close-packed environments, Ni atoms were inserted into geometrically accessible void sites near the interface. Candidate sites were identified via void analysis within a cutoff sphere, and overlaps were rejected using a hard-sphere criterion based on metallic radii.

Figure 2(c)[23)]. illustrates the resulting atomistic model of Ni/$RE$Ni$_2$ interface generated through the steps (i)-(v). Atomic coordinates are provided in Table S1.

(vi) **Full structural relaxation**

Each interface structure was relaxed using DFT, allowing both atomic positions and lattice parameters to vary. Convergence criteria and detailed settings are described in Section 2.3. This relaxation step yield (quasi-)stable interface structures for all coherent Ni/$RE$Ni$_2$ systems, enabling reliable evaluation of interfacial energies.

2.2 Interfacial energy formulation

The interfacial energy of a Ni/$RE$Ni$_2$ is defined as:

$$\text{Interfacial energy} = \frac{E_{\text{Ni}/RE\text{Ni}_2} - n_{\text{Ni}} E_{\text{Ni}}(V_*) - n_{RE\text{Ni}_2} E_{RE\text{Ni}_2}(V_*)}{2A_*}, \quad (1)$$

where $E_{\text{Ni}/RE\text{Ni}_2}$ is the total energy of the relaxed interface supercell, $E_M(V)$ and $n_M$ ($M = \text{Ni}, RE\text{Ni}_2$) are the reference bulk energy and the number of formula units at the supercell, $V_*$ is the supercell volume, and $A_*$ is the area of one interface plane. The denominator includes a factor of 2 because the supercell contains two equivalent interfaces. The $E_{\text{Ni}/RE\text{Ni}_2}$ and $V_*$ were obtained from the structure relaxation calculation.

Bulk reference energies were obtained from DFT energy-volume data and fitted to the Birch—Murnaghan equation of state (EoS) [26],

$$E(V) = E_0 + \frac{9V_0 B_0}{16}\eta^2\{(B_0' - 4)\eta + 2\}, \quad \eta \equiv \left(\frac{V_0}{V}\right)^{\frac{2}{3}} - 1. \quad (2)$$

Here $E_0, V_0, B_0,$ and $B_0'$ denote the equilibrium energy, equilibrium volume, bulk modulus, and its pressure derivative, respectively.

To quantify the mechanical contribution to interface stability, we define the strain energy per area,

$$\text{Strain energy} = \frac{E_M(V_*) - E_{0,M}}{A_*}, \quad (3)$$

representing the increase in bulk energy density upon homogeneous deformation from $V_0$ to $V_*$. This quantify the pure elastic (non-chemical) contribution of the interfacial energetics.

2.3 Computational method

All calculations were performed within the ultrasoft pseudo-potential plane-wave formalism implemented in Quantum ESPRESSO (PWSCF) [27),28]. Exchange–correlation effects were treated with the modified Pardew—Burke—Ernzerhof generalized-gradient approximation (PBEsol)[29]. Pseudo-potentials were taken from the PSlibrary[30]. For lanthanoid elements, 4f electrons were treated within the frozen-core approximation assuming a $RE^{3+}$ valence state. The plane-wave kinetic-energy cutoff was set to 90 Ry and the charge cutoff was set to 720 Ry. Brillouin-zone integrations employed Monkhorst–Pack meshes of 8 × 8 × 8 for bulk structures and 8 × 8 × 1 for interface supercells.

# 3. Results and discussion

3.1 Equation of state fitting for base crystals.

The equilibrium energies and volumes of the base crystals were obtained by fitting DFT-calculated energy–volume data to Eq. (2). Figure 3 shows the resulting fitting curves for fcc-Ni and the series of $RE$Ni$_2$ compounds ($RE$ = Sc, Y, Nd, Gd, Dy, and Lu), where the total energy is plotted as a function of the cell volume relative to the equilibrium energy. The corresponding fitting parameters are summarized in Table I.

All systems exhibit a strongly asymmetric dependence of total energy on volume. Compression below the equilibrium volume causes a steep increase in energy, reflecting a large elastic penalty associated with reducing interatomic distances. In contrast, expansion leads to a more gradual energy increase, indicating that tensile deformation is energetically more favorable.

Among the materials examined, fcc-Ni displays the steepest curvature in its energy–volume relation, confirming that Ni is elastically stiffer than the $RE$Ni$_2$ compounds. This trend is consistent with the bulk modulus $B_0$ values listed in in Table I, where $B_0$ for Ni significantly exceeds those of $RE$Ni$_2$. These results imply that, under epitaxial constraints, Ni will resist volume deformation more strongly than $RE$Ni$_2$.

3.2 Interfacial energy and lattice mismatch

Interfacial energies were evaluated using Eq. (1), and summarized in Figure 4(a). Ni/DyNi$_2$ exhibits significantly lower interfacial energy than NdNi$_2$, indicating superior interfacial stability. This trend is consistent with experimental observations of epitaxial DyNi$_2$ growth and the absence of epitaxial NdNi$_2$ growth. Strain energies of the corresponding $RE$Ni$_2$ bulk phases (Figure 4(b)) further show that DyNi$_2$ accommodates smaller strain energy than NdNi$_2$, supporting the dominant role of elasticity.

Comparison of initial (unrelaxed) and relaxed configurations reveals that systems with large initial strain energy release more strain energy upon relaxation. This indicates that elastic relaxation is a major mechanism governing the final interfacial configuration.

To quantify the elastic contribution, we analyzed lattice mismatch, defined as

$$\text{Lattice mismatch (\%)} = \frac{a_{RE\text{Ni}_2} - 2a_{\text{Ni}}}{2a_{\text{Ni}}} \times 100. \quad (4)$$

Figure 5(a) shows that interfacial energy scales linearly with the absolute value of lattice mismatch. LuNi$_2$ and DyNi$_2$, which exhibit small mismatch, show low interfacial energies, while NdNi$_2$, with larger mismatch, shows high interfacial energy.

Importantly, this trend is consistent with experimental observations, where DyNi$_2$ can grow epitaxially on Ni, while NdNi$_2$ cannot. The agreement between first-principles calculations and experimental growth behavior supports the validity of the present interfacial energy evaluation.

Figure 5(b) compares interfacial energy with strain energy and reveals that both quantities display nearly identical elemental dependence for systems with positive lattice mismatch. This confirms that elastic strain is the primary factor governing interfacial stability in Ni/*RE*Ni$_2$ systems rather than chemical bonding at interfaces.

An exception is ScNi$_2$, which shows negative lattice mismatch. Despite its low strain energy, ScNi$_2$ yields a relatively high interfacial energy. This behavior is attributed to the large bulk modulus of Ni: relaxation induces compressive strain primarily on the Ni side, creating a substantial elastic penalty that offsets the low strain energy in ScNi$_2$.

### 3.3 Prediction of interfacial energy

To generalize the relationship between lattice mismatch and interfacial energy, a linear regression model was developed using the absolute lattice mismatch as the descriptor. It also allows us to mitigate the systematic overestimation/underestimation in lattice constant due to the choice of exchange-correlation functional in DFT. The lattice constants for *RE*= Sc, Y, La, Ce, Pr, Nd, Pm, Sm, Eu, Gd, Tb, Dy, Ho, Er, Tm, Yb, and Lu were compiled from the International Centre for Diffraction Data (ICDD)[31], Crystallography Open Database (COD)[32], and this study (Table S2). Lattice constants for some compounds, such as LaNi$_2$, PmNi$_2$, and EuNi$_2$, were not available in certain databases. The atomic radii of RE were obtained from database[33], and both the lattice mismatch and atomic radius were plotted in Figure 6. The obtained lattice mismatch calculated in this study were close to the values calculated from lattice parameters in the database. The lattice mismatch decreased as the atomic radius of the RE elements decreased, with minor database dependent variations.

A least-squares fit using the DFT-computed interfacial energies yielded a regression model with $R^2 = 0.88$, as shown in Figure 7(a). demonstrating that a simple linear relationship captures the essential dependence of interfacial energy on lattice mismatch.

Predicted interfacial energies are summarized in Figure 7(b). Systems predicted to have lower interfacial energies than DyNi$_2$—such as HoNi$_2$, ErNi$_2$, TmNi$_2$, YbNi$_2$, and LuNi$_2$—are expected to form stable interfaces favorable for epitaxial growth. Conversely, LaNi$_2$ and PrNi$_2$ exhibit higher predicted interfacial energies than NdNi$_2$ and are thus expected to produce unstable interfaces. Indeed, LaNi$_2$ does not appear as a stable phase in the phase

diagram[34]). Interestingly, CeNi$_2$ and PrNi$_2$ exhibit relatively small lattice mismatches compared to their large atomic radii of RE elements, resulting in CeNi$_2$ having lower interfacial energy than NdNi$_2$.

Although interfacial energy provides a strong indicator of epitaxial stability, it may not serve as a strict criterion for epitaxial growth. Further investigation is required to fully assess the stability of all compounds by considering additional factors such as diffusion kinetics and finite-temperature effects.

## 4. Conclusions

In this study, we systematically investigated the interfacial stability of Ni/*RE*Ni$_2$ interfaces using first-principles calculations, with emphasis on the role of lattice mismatch and elastic strain.

The equation-of-state analysis revealed an asymmetric energy–volume dependence for all systems: compressive deformation incurs a much higher energetic penalty than tensile deformation. Furthermore, fcc-Ni was found to be elastically stiffer than *RE*Ni$_2$ compounds, indicating that compressive strain on the Ni side can significantly increase the interfacial energy.

Explicit interfacial-energy calculations showed a clear correlation between lattice mismatch and stability: smaller lattice mismatch leads to lower interfacial energy and, consequently, more stable interfaces. Comparison with strain energies demonstrated that the observed elemental dependence is predominantly governed by elastic strain rather than chemical bonding effects. In particular, instability in Ni/*RE*Ni$_2$ interfaces originates mainly from strain accumulation associated with compressive deformation of the *RE*Ni$_2$. An exception was identified for ScNi$_2$, where relaxation introduces compressive strain primarily in the Ni region due to the high stiffness of fcc-Ni, resulting in an increased interfacial energy despite the small strain energy of ScNi$_2$.

Based on these insights, we developed a linear regression model using the absolute lattice mismatch as a descriptor, achieving a high predictive accuracy ($R^2 = 0.88$). This model allowed qualitative prediction of interfacial energies for a broad range of RE elements and provides a practical guideline for assessing interfacial stability and the likelihood of epitaxial growth.

Overall, our results indicate that interface stability in Ni/*RE*Ni$_2$ systems is largely dictated by elastic strain. The framework established here offers a transferable approach for evaluating interface energetics and can aid in analyzing stability of coherent heterostructures.

Future studies in conjunction with experimental validation incorporating finite-temperature effects, interfacial diffusion, and more complex interface geometries will further refine the predictive capability of this approach and its connection to experimentally observed growth behavior.

## Acknowledgments

We would thank Yoyo Hinuma in National Institute of Advanced Industrial Science and Technology for helpful guidance in using Crystal Model Processing Software.

This work was supported by JSPS KAKENHI Grant Number 22K14524.

This research used computational resources of Wisteria/BDEC-01 Odyssey (the University of Tokyo).

## References

1) T. Oishi, H. Konishi, T. Nohira, M. Tanaka and T. Usui, Kagaku Kogaku Ronbunshu **36** [4], 299 (2010).
2) T. Oishi, M. Yaguchi, Y. Katasho, H. Konishi and T. Nohira, J. Electrochem. Soc. **167** [16], 163505 (2020).
3) T. Oishi, M. Yaguchi, Y. Katasho and T. Nohira, J. Electrochem. Soc. **168** [10], 103504 (2021).
4) T. Nohira, in *Molten Salts Chemistry* (Elsevier, 2013) pp. 287.
5) W. Han, M. Li, M. L. Zhang and Y. D. Yan, Rare Met. **35** [11], 811 (2016).
6) H. Konishi, T. Nohira and Y. Ito, J. Electrochem. Soc. **148** [7], C506 (2002).
7) H. Konishi, T. Nohira and Y. Ito, Electrochim. Acta **48** [5], 563 (2003).
8) Y. Katasho and T. Oishi, Electrochem. Commun. **138**, 107287 (2022).
9) Y. Katasho and T. Oishi, Electrochemistry **92** [4], 23 (2024).
10) H. Konishi, H. Ono, T. Nohira and T. Oishi, ECS Trans. **50** [11], 463 (2013).
11) S. Wang, H. Wang, B. Ma, B. Zhu, Y. Yang, M. Qiu and M. Zhang, Mater. Chem. Phys. **339**, 130805 (2025).
12) Y. Katasho and T. Oishi, J. Electrochem. Soc. **172** [6], 062505 (2025).
13) W. Han, Q. Zhao, J. Wang, M. Li, W. Liu, M. Zhang, X. Yang and Y. Sun, J. Rare Earths **35** [1], 90 (2017).
14) H. Konishi, Y. Yoshihara, H. Ono, T. Usui, T. Oishi and T. Nohira, J. Jpn. Soc. Exp. Mech. **10**, s215 (2010).
15) S. Wang, M. Li, W. Han, M. Zhang, W. Wang, W. Wang, X. Yang and Y. Sun, J. Alloys Compd. **777**, 1211 (2019).
16) T. Nohira, H. Kambara, K. Amezawa and Y. Ito, J. Electrochem. Soc. **152** [4], C183 (2005).
17) T. Yin, Y. Liang, J. Qu, P. Li, R. An, Y. Xue, M. Zhang, W. Han, G. Wang and Y. Yan, J. Electrochem. Soc. **164** [13], D835 (2017).
18) T. Iida, T. Nohira and Y. Ito, Electrochim. Acta **46** [16], 2537 (2001).
19) H. Konishi, K. Mizuma, H. Ono, E. Takeuchi, T. Nohira and T. Oishi, ECS Trans. **50** [11], 561 (2013).
20) M. Li, J. Wang, W. Han, Y. Dong, W. Wang, M. Zhang, X. Yang and Y. Sun, J.

Electrochem. Soc. **165** [14], D704 (2018).
21) T. Iida, T. Nohira and Y. Ito, Electrochim. Acta **48** [11], 1531 (2003).
22) J. Li, J. Wang, P. He, J. Shu, J. Yang, W. Li, X. Yin, D. Zeng, Q. Wu and Y. Wei, *1 Morphological Impact of Ni-Based Electrodes (Solid Ni Sheets vs. 2 Ni(II)-Assisted W Meshes) on Lu(III) Electrochemical Behavior and 3 Efficient Extraction Mechanism in LiCl-KCl Molten Salts* (SSRN: https://ssrn.com/abstract=5941710).
23) K. Momma and F. Izumi, J. Appl. Crystallogr. **44** [6], 1272 (2011).
24) Y. Hinuma, T. Kamachi and N. Hamamoto, Mater. Trans. **61** [1], 78 (2020).
25) Y. Hinuma, J. Comput. Chem. Jpn. **23** [1], A2 (2024).
26) D. S. Sholl and J. A. Steckel, *Density Functional Theory: A Practical Introduction* (Wiley, 2009) 1st edn.
27) P. Giannozzi, S. Baroni, N. Bonini, M. Calandra, R. Car, C. Cavazzoni, D. Ceresoli, G. L. Chiarotti, M. Cococcioni, I. Dabo, A. Dal Corso, S. De Gironcoli, S. Fabris, G. Fratesi, R. Gebauer, U. Gerstmann, C. Gougoussis, A. Kokalj, M. Lazzeri, L. Martin-Samos, N. Marzari, F. Mauri, R. Mazzarello, S. Paolini, A. Pasquarello, L. Paulatto, C. Sbraccia, S. Scandolo, G. Sclauzero, A. P. Seitsonen, A. Smogunov, P. Umari and R. M. Wentzcovitch, J. Phys.: Condens. Matter **21** [39], 395502 (2009).
28) P. Giannozzi, O. Andreussi, T. Brumme, O. Bunau, M. Buongiorno Nardelli, M. Calandra, R. Car, C. Cavazzoni, D. Ceresoli, M. Cococcioni, N. Colonna, I. Carnimeo, A. Dal Corso, S. De Gironcoli, P. Delugas, R. A. DiStasio, A. Ferretti, A. Floris, G. Fratesi, G. Fugallo, R. Gebauer, U. Gerstmann, F. Giustino, T. Gorni, J. Jia, M. Kawamura, H.-Y. Ko, A. Kokalj, E. Küçükbenli, M. Lazzeri, M. Marsili, N. Marzari, F. Mauri, N. L. Nguyen, H.-V. Nguyen, A. Otero-de-la-Roza, L. Paulatto, S. Poncé, D. Rocca, R. Sabatini, B. Santra, M. Schlipf, A. P. Seitsonen, A. Smogunov, I. Timrov, T. Thonhauser, P. Umari, N. Vast, X. Wu and S. Baroni, J. Phys.: Condens. Matter **29** [46], 465901 (2017).
29) J. P. Perdew, A. Ruzsinszky, G. I. Csonka, O. A. Vydrov, G. E. Scuseria, L. A. Constantin, X. Zhou and K. Burke, Phys. Rev. Lett. **100** [13], 136406 (2008).
30) A. Dal Corso, Comput. Mater. Sci. **95**, 337 (2014).
31) International Centre for Diffraction Data (ICDD), *PDF-2 (Powder Diffraction File)* (2014).
32) S. Gražulis, D. Chateigner, R. T. Downs, A. F. T. Yokochi, M. Quirós, L. Lutterotti, E. Manakova, J. Butkus, P. Moeck and A. Le Bail, J. Appl. Crystallogr. **42** [4], 726 (2009).
33) Royal Society of Chemistry, Periodic Table, https://periodic-table.rsc.org/, (accessed 7 January 2026).
34) H. Okamoto, J. Phase Equilib. Diffus. **41** [5], 722 (2020).

## Figures

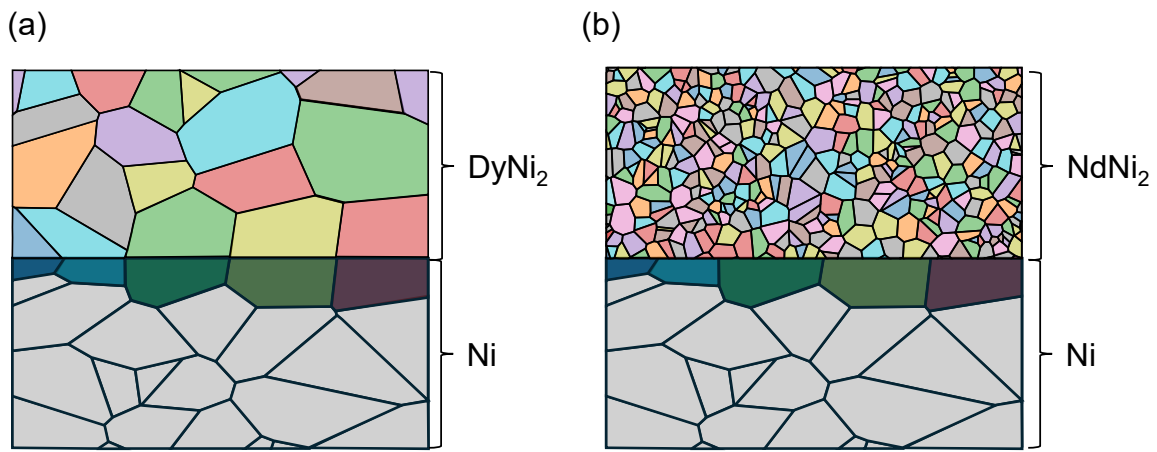

**Figure 1.** (a) Schematic illustration of the crystal grain structure of Ni electrode after electrochemical alloying in molten LiCl–KCl–DyCl$_3$ and (b) in molten LiCl–KCl–NdCl$_3$.

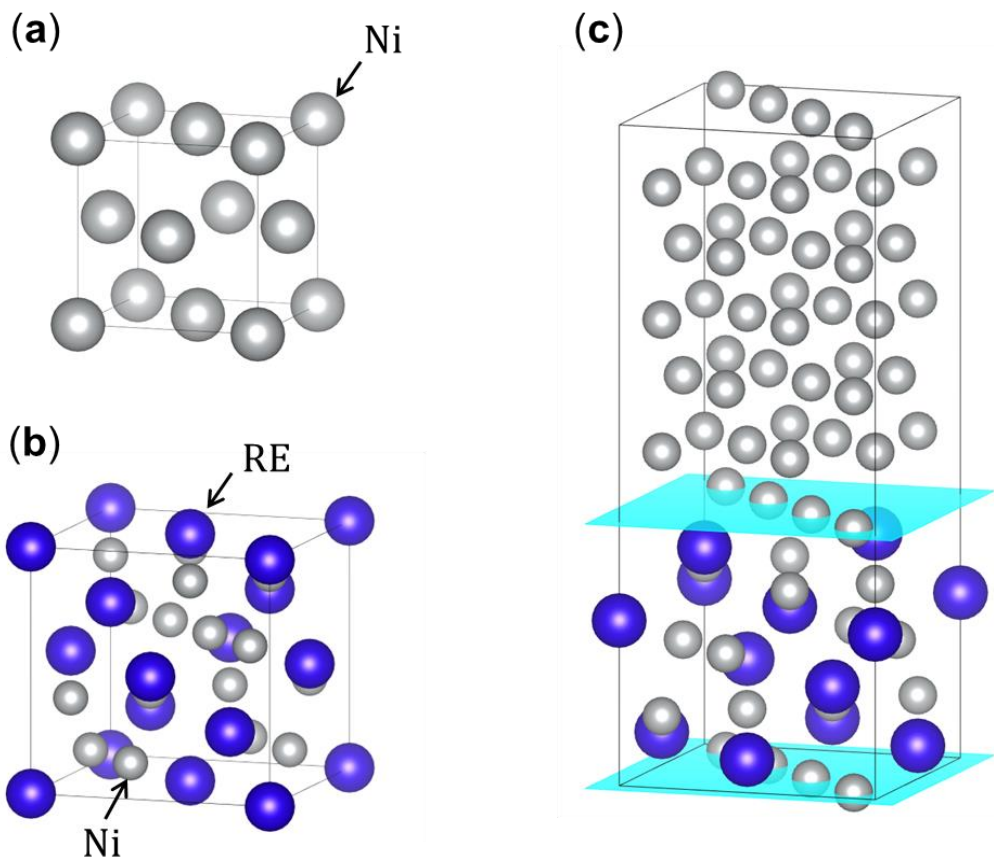

**Figure 2.** Unit cell of (a) fcc-Ni, (b) $RE$Ni$_2$, and (c) the interface model of Ni/$RE$Ni$_2$ which is illustrated by VESTA package [23)].

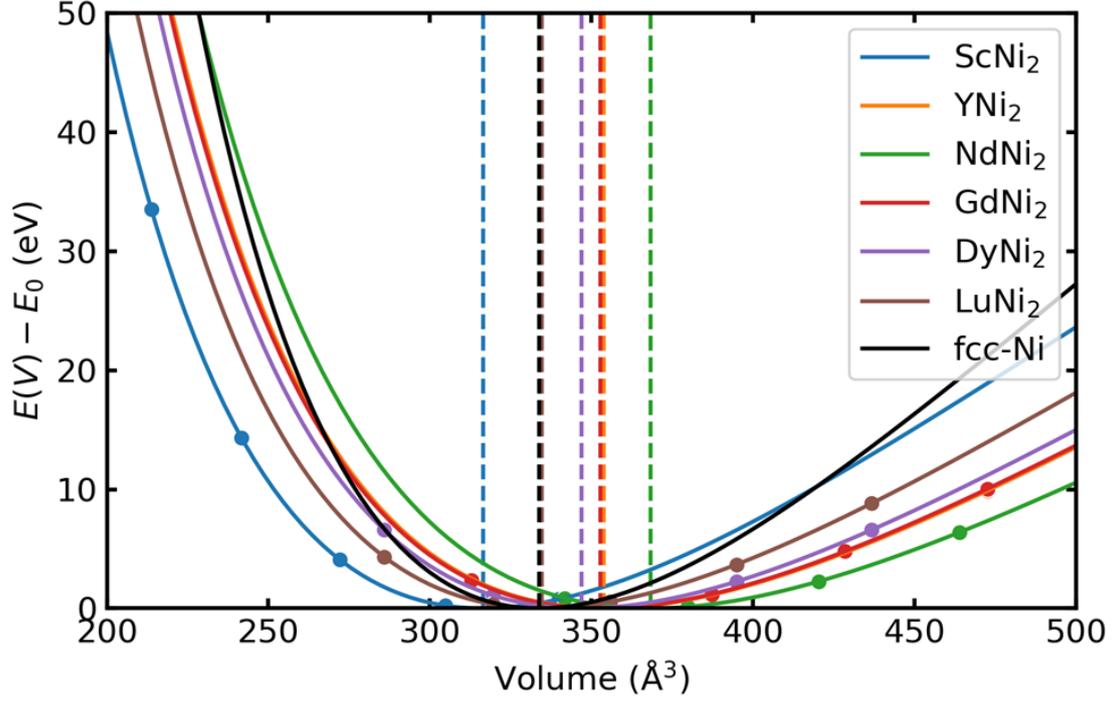

**Figure 3.** Volume-energy curve for the base crystals. Solid circles, solid lines, and dashed vertical lines represent the reference (computed) values, the fitting curves, and the values of equilibrium volume, respectively. For fcc-Ni, the values are estimated from a (2×2×2) supercell.

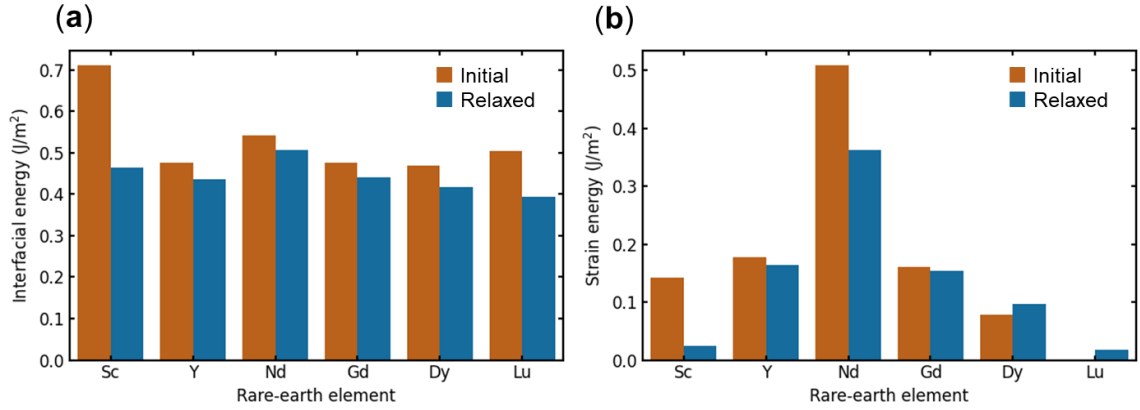

**Figure 4.** (a) Interfacial energy of Ni/$RE$Ni$_2$ interface and (b) strain energy of base $RE$Ni$_2$ crystal ($RE$ = Sc, Y, Nd, Gd, Dy, and Lu). "Initial" (orange bar) and "Relaxed" (blue bar) represent the energies on the initial structures (before relaxation) and those of the relaxed structures, respectively.

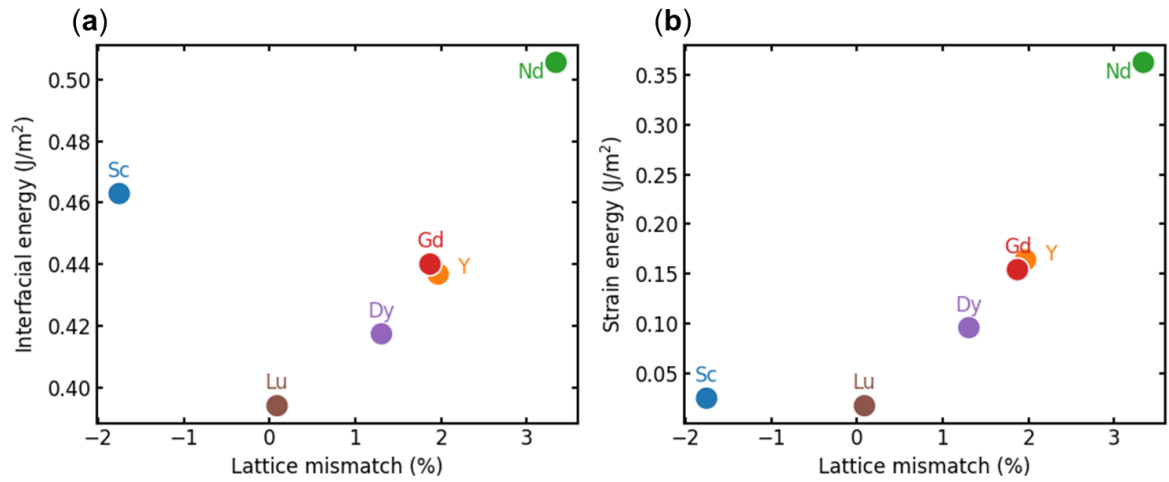

**Figure 5.** Relationship between lattice mismatch and (a) interfacial energy of Ni/$RE$Ni$_2$ interface, or (b) strain energy of base $RE$Ni$_2$ crystal ($RE$ = Sc, Y, Nd, Gd, Dy, and Lu).

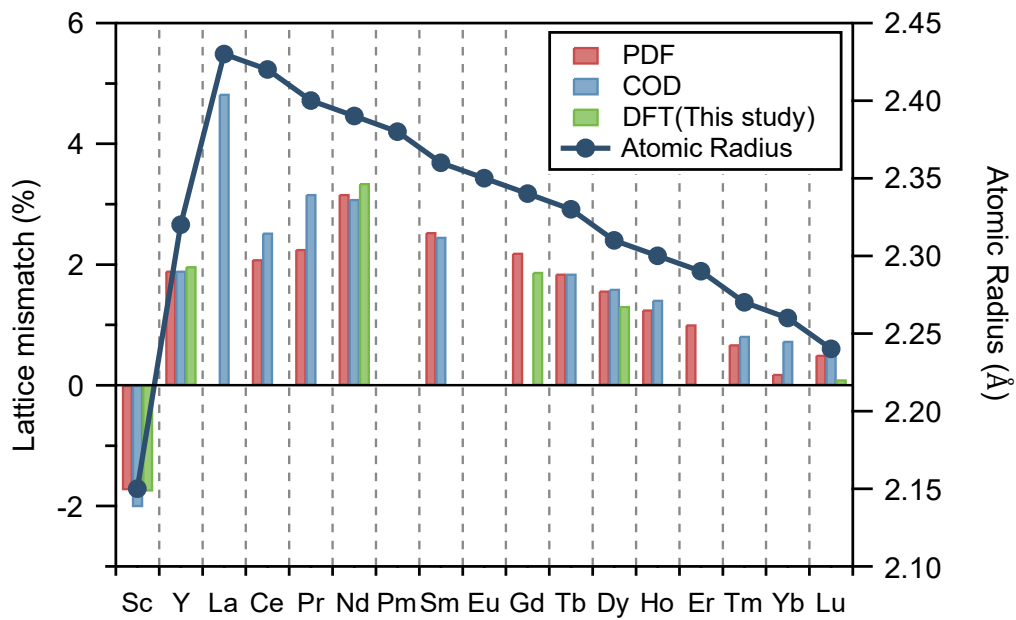

**Figure 6.** Lattice mismatch between Ni and each $RE$Ni$_2$ compound, and the atomic radius of the $RE$ elements obtained from International Centre for Diffraction Data (ICDD), Crystallography Open Database (COD), and this study (DFT).

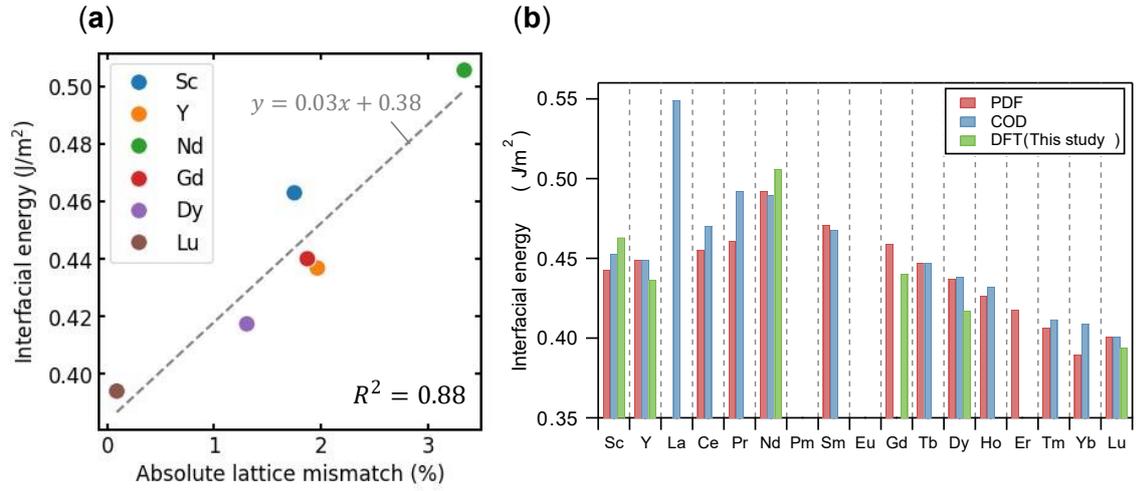

**Figure 7.** (a) Linear regression model to predict interfacial energies from absolute values of lattice mismatch. (b) Prediction results obtained by referring the literature values.

**Table 1.** Equation of state fitting parameters for base crystals.

|  | RE | $V_0$ (Å³) | $E_0$ (eV) | $B_0$ (eV/Å³) | $B_0'$ |
|---|---|---|---|---|---|
| Ni (2×2×2) | - | 333.75 | -43640 | 1.44 | 4.81 |
| ScNi$_2$ | Sc | 316.51 | -32474 | 1.00 | 4.32 |
| YNi$_2$ | Y | 353.82 | -30914 | 0.840 | 4.44 |
| DyNi$_2$ | Dy | 347.00 | -35400 | 0.861 | 4.42 |
| NdNi$_2$ | Nd | 368.33 | -35435 | 0.786 | 4.49 |
| GdNi$_2$ | Gd | 352.85 | -35304 | 0.841 | 4.45 |
| LuNi$_2$ | Lu | 334.64 | -35414 | 0.916 | 4.40 |

Supplementary data

**Table S1.** Configuration of Ni/*RE*Ni$_2$ interface model. Positions (x, y, z) are represented in the fractional coordinate.

| Species | x | y | z | Species | x | y | z |
|---|---|---|---|---|---|---|---|
| RE | 0 | 0.5 | 0.05 | Ni | 0.375 | 0.625 | 0.5 |
| RE | 0.5 | 0 | 0.05 | Ni | 0.625 | 0.875 | 0.5 |
| RE | 0.75 | 0.75 | 0.15 | Ni | 0.875 | 0.375 | 0.6 |
| RE | 0.25 | 0.25 | 0.15 | Ni | 0.125 | 0.125 | 0.6 |
| RE | 0.5 | 0.5 | 0.25 | Ni | 0.875 | 0.875 | 0.6 |
| RE | 0 | 0 | 0.25 | Ni | 0.125 | 0.625 | 0.6 |
| RE | 0.25 | 0.75 | 0.35 | Ni | 0.375 | 0.375 | 0.6 |
| RE | 0.75 | 0.25 | 0.35 | Ni | 0.625 | 0.125 | 0.6 |
| Ni | 0.625 | 0.625 | 0 | Ni | 0.375 | 0.875 | 0.6 |
| Ni | 0.375 | 0.375 | 0 | Ni | 0.625 | 0.625 | 0.6 |
| Ni | 0.375 | 0.625 | 0.1 | Ni | 0.875 | 0.125 | 0.7 |
| Ni | 0.875 | 0.125 | 0.1 | Ni | 0.125 | 0.375 | 0.7 |
| Ni | 0.625 | 0.375 | 0.1 | Ni | 0.875 | 0.625 | 0.7 |
| Ni | 0.125 | 0.875 | 0.1 | Ni | 0.125 | 0.875 | 0.7 |
| Ni | 0.125 | 0.625 | 0.2 | Ni | 0.375 | 0.125 | 0.7 |
| Ni | 0.625 | 0.125 | 0.2 | Ni | 0.625 | 0.375 | 0.7 |
| Ni | 0.875 | 0.375 | 0.2 | Ni | 0.375 | 0.625 | 0.7 |
| Ni | 0.375 | 0.875 | 0.2 | Ni | 0.625 | 0.875 | 0.7 |
| Ni | 0.875 | 0.625 | 0.3 | Ni | 0.875 | 0.375 | 0.8 |
| Ni | 0.375 | 0.125 | 0.3 | Ni | 0.125 | 0.125 | 0.8 |
| Ni | 0.125 | 0.375 | 0.3 | Ni | 0.875 | 0.875 | 0.8 |
| Ni | 0.625 | 0.875 | 0.3 | Ni | 0.125 | 0.625 | 0.8 |
| Ni | 0.125 | 0.125 | 0.4 | Ni | 0.375 | 0.375 | 0.8 |
| Ni | 0.375 | 0.375 | 0.4 | Ni | 0.625 | 0.125 | 0.8 |
| Ni | 0.625 | 0.625 | 0.4 | Ni | 0.375 | 0.875 | 0.8 |
| Ni | 0.875 | 0.875 | 0.4 | Ni | 0.625 | 0.625 | 0.8 |
| Ni | 0.875 | 0.875 | 0 | Ni | 0.875 | 0.125 | 0.9 |
| Ni | 0.125 | 0.125 | 0 | Ni | 0.125 | 0.375 | 0.9 |
| Ni | 0.875 | 0.125 | 0.5 | Ni | 0.875 | 0.625 | 0.9 |
| Ni | 0.125 | 0.375 | 0.5 | Ni | 0.125 | 0.875 | 0.9 |
| Ni | 0.875 | 0.625 | 0.5 | Ni | 0.375 | 0.125 | 0.9 |
| Ni | 0.125 | 0.875 | 0.5 | Ni | 0.625 | 0.375 | 0.9 |
| Ni | 0.375 | 0.125 | 0.5 | Ni | 0.375 | 0.625 | 0.9 |
| Ni | 0.625 | 0.375 | 0.5 | Ni | 0.625 | 0.875 | 0.9 |

**Table S2** Lattice mismatch between Ni and each $RE$Ni$_2$ compound, and the atomic radius of the RE elements.

| Compound | RE | Atomic radius of RE (Å) | PDF | | | COD | | | This study | |
|---|---|---|---|---|---|---|---|---|---|---|
| | | | Lattice constant (Å) | Lattice mismatch to Ni (%) | PDF number | Lattice constant (Å) | Lattice mismatch to Ni (%) | COD ID | Lattice constant (Å) | Lattice mismatch to Ni (%) |
| Ni | | — | 3.524 | — | 00-004-0850 | 3.524 | — | 9008476 | 3.468 | — |
| ScNi$_2$ | Sc | 2.15 | 6.926 | -1.725 | 03-065-5763 | 6.906 | -2.009 | 1522485 | 6.815 | -1.752 |
| YNi$_2$ | Y | 2.32 | 7.181 | 1.893 | 03-065-3043 | 7.181 | 1.893 | 1538213 | 7.073 | 1.966 |
| LaNi$_2$ | La | 2.43 | — | — | — | 7.387 | 4.816 | 1537960 | — | — |
| CeNi$_2$ | Ce | 2.42 | 7.194 | 2.077 | 03-065-0105 | 7.225 | 2.517 | 1524647 | — | — |
| PrNi$_2$ | Pr | 2.40 | 7.206 | 2.248 | 03-065-3046 | 7.27* | 3.156 | 1521762 | — | — |
| NdNi$_2$ | Nd | 2.39 | 7.27* | 3.156 | 03-065-4584 | 7.265 | 3.085 | 1522930 | 7.168 | 3.341 |
| PmNi$_2$ | Pm | 2.38 | — | — | — | — | — | — | — | — |
| SmNi$_2$ | Sm | 2.36 | 7.226 | 2.531 | 03-065-5536 | 7.22* | 2.446 | 1522577 | — | — |
| EuNi$_2$ | Eu | 2.35 | — | — | — | — | — | — | — | — |
| GdNi$_2$ | Gd | 2.34 | 7.202 | 2.191 | 03-065-6163 | — | — | — | 7.066 | 1.873 |
| TbNi$_2$ | Tb | 2.33 | 7.178 | 1.843 | 00-038-1472 | 7.178 | 1.845 | 1538705 | — | — |
| DyNi$_2$ | Dy | 2.31 | 7.157 | 1.558 | 00-050-1533 | 7.16* | 1.595 | 1524887 | 7.027 | 1.306 |
| HoNi$_2$ | Ho | 2.30 | 7.136 | 1.254 | 03-065-5543 | 7.147 | 1.410 | 1522472 | — | — |
| ErNi$_2$ | Er | 2.29 | 7.118 | 0.999 | 03-065-5800 | — | — | — | — | — |
| TmNi$_2$ | Tm | 2.27 | 7.095 | 0.673 | 03-065-6748 | 7.105 | 0.814 | 1523330 | — | — |
| YbNi$_2$ | Yb | 2.26 | 7.06* | 0.176 | 03-065-5017 | 7.099 | 0.729 | 1538937 | — | — |
| LuNi$_2$ | Lu | 2.24 | 7.083 | 0.502 | 03-065-9356 | 7.083 | 0.502 | 1522270 | 6.943 | 0.089 |

*Values from the database are given with a precision of 0.01 Å.